\documentclass[12pt]{article}
\usepackage{amsmath}
\usepackage{amsfonts}
\usepackage{mathrsfs}
\usepackage{hyperref}
\allowdisplaybreaks[1]
\title{\bf Massive mixed symmetry field dynamics \\ in open bosonic string theory} 
\author{V.A. Krykhtin\footnote{krykhtin@tspu.edu.ru }
\\[1em]
{\it Department of Theoretical Physics, Tomsk State Pedagogical  University,}\\ {\it  634061, Tomsk,  Russia }
}

\date{}

\begin{document}

\maketitle

\begin{abstract}
We consider the sigma-model description of an open string interacting with massive fields of the fourth (third massive) level.
Equations of motion for the background fields are obtained by demanding that the renormalized
operator of the energy-momentum tensor trace vanishes.
\end{abstract}
\section{\bf Introduction}\setcounter{equation}{0}
As is known, in higher dimensions $d>4$
the totally symmetric tensor fields are not enough
to cover all the irreducible representations of the Poincare group (see e.g. \cite{Vasiliev:2004qz,Rahman:2015pzl}) 
and we should take into account the fields with mixed symmetry of the indices as well. 
Therefore a higher spin field theory in higher dimensions should describe a mixed symmetry fields dynamics
(see e.g. \cite{Alkalaev:2003qv,Alkalaev:2005kw,Buchbinder:2007ix,Buchbinder:2008kw,Buchbinder:2009pa,Skvortsov:2008sh,Skvortsov:2010nh,Campoleoni:2008jq,Campoleoni:2009gs}).
Mixed symmetry fields naturally arise in context of superstring theory
(see more about relation the string and the higher spin theories e.g. in 
\cite{Rahman:2015pzl,Buchbinder:1999ar,Sagnotti:2010at,Sagnotti:2013bha,Lee:2017utr}).
One of the possibilities to derive equations of motion on higher spin fields is to consider
a string theory in background fields.

The theory of a string in massless background fields provides a possibility
to consider string interactions in the low energy approximation \cite{1,2,3,4}
(see reviews \cite{5,6}).
The quantum  Weyl invariance of the theory is a crucial point in this approach. 
It means that the renormalized operator of the energy-momentum tensor trace must vanish
resulting in equations of motion for background fields. The general
analysis of the renormalized operator for a bosonic string interacting
with background metric, antisymmetric tensor and dilaton was performed
in refs.\cite{7,8} (see also review \cite{5}).

A natural development of the approach is to consider 
string in background fields which are connected with massive modes in
the string spectrum. 
A string
interacting with any finite number of massive background fields is
non-renormalizable theory and requires infinite number of
countertermes. So we have to deal with the theory containing infinite
number of terms in classical action which describe interaction with
background fields of all the massive modes. The only massive field that
does not require infinite number of counterterms is the field of
tachyon but in this case non-perturbative effects play a crucial role
\cite{9,10,11,12,13} (see also review \cite{14}).

Several attempts were undertaken to describe string in massive
background fields \cite{12,13,14,15,16,17,18,19,20,21,22,23}. 
All these investigations (excluding work on
the tachyon problem) concerned mainly only linear equations of motion for
background fields. The linear approximation is of great importance
since the equations for background fields in this approximation should
correspond to the known equations defining the string spectrum. It
was turned out that even at the linear level the whole clarity is
absent and the very possibility of going beyond the linear
approximation presents difficulties.

In ref.\cite{24} it was proposed an approach to the string theory in massive
background fields that represents a direct generalization of the
$\sigma$-model approach to the string theory in massless background fields.
there it was shown that for a closed string interacting with background fields of all the massive
modes the renormalization has a special structure. Namely,
the renormalization of background fields of any massive level requires
consideration only of background fields of this level and of all the lower
ones, but is not affected by the infinite number of background fields
belonging to higher levels. In principle, our approach allows to go beyond
the linear approximation.

 Further in \cite{Buchbinder:1994rk,Buchbinder:1996vq} we developed the approach \cite{24}
to its application to the theory of open string in massive background
fields.  
There we investigated the renormalization in the theory of a string interacting with arbitrary
background fields both on the world sheet and its boundary
and  that the renormalization has the same structure as in ref.\cite{24}.
The theory of an open string is a field theory in space-time
with boundary.  Various aspects of quantum calculations in the theory
of open strings were discussed in ref. \cite{29}. As was noted in
pioneer works \cite{2}, interaction of an open string with background
fields corresponding to the open string spectrum is completely
concentrated at the boundary of the world sheet. Detailed investigation
of open string in massless background fields was conducted in
refs. \cite{26,27,28}. Questions of quantum field theory on a manifold with
boundary were studied in works \cite{30,31}.

Our aim consists in construction of a $\sigma$-model type action describing
interaction of an open string with massive background fields of the fourth level and in deriving
from the quantum Weyl invariance condition effective equations of motion
for these fields. So we have to build up the renormalized operator of the
energy-momentum tensor trace and to demand that it vanishes.
To construct the renormalized operator of the energy-momentum tensor trace
one has to renormalize both the background fields and the
composite operators. Renormalization of fields is
constructed by demanding that divergences of the quantum effective action
vanish.
Since we are interested in linear
approximation it is suffitient to restrict ourselves to the study of
divergences generated by fields of the only given massive level  \cite{24,Buchbinder:1994rk,Buchbinder:1996vq}.
Contributions of all other levels contain products of several background
fields and so are beyond the linear approximation.

The paper is organized as follows. Section 2  we consider sigma-model description of an open string interacting with
massive fields of the fourth (third massive) level, describe its symmetries 
and construct one-loop renormalization of the background fields. 
The renormalization of composite operators defining the energy-momentum tensor trace is carried
out in section 3 and in section 4 linear equations of motion for background fields are derived and discussed.

\section{\bf Action and renormalization of background fields}\setcounter{equation}{0}

In this section we will consider an open string interacting with all the
massive fields of the fourth (third massive) level, describe its symmetries and construct
one-loop renormalization of background fields. The action should be the sum
of the free string action $S_0$ and the action $S_I$ describing interaction
with the fields of the fourth massive level and containing all possible
terms of the fourth order in derivatives.
After rescaling string coordinates $x^{\mu}\to\sqrt{\alpha'}x^{\mu}$ the classical action has the form
\begin{eqnarray}
S[x]&=&S_0[x]+S_I[x]
\label{300}
\\
S_0[x]
&=&
\frac{1}{4\pi}\int_{M^2}d^2z \;\partial_a x^\mu \, \partial^a x_\mu
\label{3.0}
\\
S_I[x]
&=&
\frac{\alpha^{\prime3/2}}{2\pi}\int_{\partial M}dt \,e\,\Bigl\{
\dot{x}^\mu\dot{x}^\nu\dot{x}^\rho\dot{x}^\sigma A_{\mu\nu\rho\sigma}
+\dot{x}^\mu\dot{x}^\nu\ddot{x}^\rho B_{\mu\nu\rho}
\nonumber
\\
&&\hspace*{17ex}{}
+\ddot{x}^\mu\ddot{x}^\nu C_{\mu\nu}
+\dot{x}^\mu\dddot{x}^\nu D_{\mu\nu}
+\ddddot{x}^\mu E_{\mu}
\Bigr\}
\label{3.1}
\\[0.3em]
&&
A_{\mu\nu\rho\sigma}=A_{(\mu\nu\rho\sigma)}
\qquad
B_{\mu\nu\rho}=B_{(\mu\nu)\rho}
\qquad
C_{\mu\nu}=C_{(\mu\nu)}\nonumber
\end{eqnarray}
The integral in
(\ref{3.0}) is taken over the whole world sheet whicn is assumed to be an uppere half-plane $y\ge0$, $z=(t,y)$ and in
(\ref{3.1}) is taken over the boundary of the world sheet, and we use the euclidian
metrics.
Possibility to add an arbitrary total derivative to the Lagrangian in
(\ref{3.1}) yields symmetry of the theory under transformations
\begin{align}
&\delta A_{\mu\nu\rho\sigma}=\partial_{(\rho}\Lambda_{\mu\nu\sigma)}
&&\delta B_{\mu\nu\rho}=3\Lambda_{\mu\nu\rho}+\partial_{(\mu}\Lambda_{\nu)\rho}
&&\delta C_{\mu\nu}=\Lambda_{(\mu\nu)}
\nonumber
\\
&\delta D_{\mu\nu}=\Lambda_{\mu\nu}+\partial_\mu \Lambda_\nu
&&\delta E_{\mu}=\Lambda_{\mu}
\label{3.2}
&&\Lambda_{\rho\mu\nu}=\Lambda_{(\rho\mu\nu)}
\end{align}
where $\Lambda_{\rho\mu\nu}(x)=\Lambda_{(\rho\mu\nu)}(x)$, $\Lambda_{\mu\nu}(x)$ and $\Lambda_{\mu}(x)$ are arbitrary functions playing the
role of transformation parameters. As one can see the fields $D_{\mu\nu}(x)$ and
$E_\mu(x)$ are Stuckelberg ones and the symmetry allows us to choose them to be equal to zero. 
Also we can remove the completely symmetric part of the field $B_{\mu\mu\rho}$ and index symmetry of this field will  correspond to the `hook' Young diagram. 
Thus the essential background fields are
$A_{(\mu\nu\rho\sigma)}(x)$, $B_{\mu\mu\rho}(x)$, $C_{(\mu\nu)}(x)$.

To renormalize the background fields we should first calculate divergences of the quantum effective action.
The one-loop correction to the effective action is
\begin{eqnarray}
\label{3.3}
\frac{1}{2}\mbox{Tr\,}\ln(S_{0\alpha\beta}+\frac{\alpha'^{3/2}}{2\pi}V_{\alpha\beta}),
\end{eqnarray}
where we denote
\begin{eqnarray}
S_{0\alpha\beta}\equiv\frac{\delta^2S_0[x]}{\delta x^\alpha\delta x^\beta},
\hspace{3em}
\frac{\alpha'^{3/2}}{2\pi}V_{\alpha\beta}\equiv
\frac{\delta^2S_I[x]}{\delta x^\alpha\delta x^\beta}
\end{eqnarray}
(\ref{3.3}) is expanded into series in powers of $(\alpha')^{1/2}$:
\begin{eqnarray}
\frac{1}{2}\mbox{Tr\,}\ln(S_{0\alpha\beta}+\frac{\alpha'^{3/2}}{2\pi}V_{\alpha\beta})=
\frac{1}{2}\mbox{Tr\,}\ln S_{0\alpha\beta}+\frac{\alpha'^{3/2}}{2}\mbox{Tr\,}V_{\alpha\gamma}
G^{\gamma\beta}+O(\alpha'^2),
\label{3.5}
\end{eqnarray}
where Green function $G^{\alpha\beta}$ of a free string is determined by
the equation
\begin{eqnarray}
&&
2\pi S_{0\alpha\beta}G^{\beta\gamma}=\delta^\gamma_\alpha .
\\
&&
G^{\mu\nu}=\delta^{\mu\nu}\int\limits^{+\infty}_{-\infty}\!\!
\frac{dp_1}{2\pi}\int\limits^{+\infty}_{-\infty}\!\!\frac{dp_2}{2\pi}
\,\frac{1}{p^2}\,e^{ip_1(t-t')}\left(e^{ip_2(y-y')}+e^{ip_2(y+y')}\right)
.
\end{eqnarray}

Divergences of the first term in (\ref{3.5}) are cancelled by the
corresponding contribution of the ghosts providing that D$=26$. 
Terms $O(\alpha'^2)$ do not contribute to renormalization of the background fields under consideration 
and so they are omitted.

Calculation of $V_{\alpha\beta}$ gives
\begin{eqnarray}
\label{3.9}
V_{\alpha\beta}(t,y;t',y')=\delta_{\partial M}(z)
\delta_{\partial M}(z')\sum_{k=0}^4V_{(k)\alpha\beta}
\frac{d^k\delta(t-t')}{dt^k}
\end{eqnarray}
where $z=(t,y)$, $z'=(t',y')$ and
\begin{eqnarray*}
V_{(0)\alpha\beta}
&=&
\dot{x}^\mu\dot{x}^\nu\dot{x}^\rho\dot{x}^\sigma\Bigl\{
\partial^2_{\alpha\beta}A_{\mu\nu\rho\sigma}
-4\partial^2_{\sigma\beta}A_{\mu\nu\rho\alpha}
+\partial^3_{\rho\sigma\beta}B_{\mu\nu\alpha}
-\partial^4_{\nu\rho\sigma\beta}D_{\mu\alpha}
+\partial^5_{\mu\nu\rho\sigma\beta}E_{\alpha}
\Bigr\}
\nonumber
\\
&&{}
+\dot{x}^\mu\dot{x}^\nu\ddot{x}^\rho\Bigl\{
\partial^2_{\alpha\beta}B_{\mu\nu\rho}+\partial^2_{\rho\beta}B_{\mu\nu\alpha}
+4\partial^2_{\nu\beta}B_{\rho\mu\alpha}-2\partial^2_{\nu\beta}B_{\alpha\mu\rho}
\\
&&\hspace{17.5ex}{}
-12\partial_\beta A_{\mu\nu\rho\alpha}+2\partial^3_{\mu\nu\beta}C_{\rho\alpha}
-3\partial^3_{\mu\nu\beta}D_{\rho\alpha}-3\partial^3_{\nu\rho\beta}D_{\mu\alpha}
+6\partial^4_{\mu\nu\rho\beta}E_{\alpha}
\Bigr\}
\nonumber
\\
&&{}
+\ddot{x}^\mu\ddot{x}^\nu\Bigl\{
\partial^2_{\alpha\beta}C_{\mu\nu}
+2\partial^2_{\nu\beta}C_{\mu\alpha}
+2\partial_{\beta}B_{\mu\nu\alpha}
-2\partial_{\beta}B_{\alpha\mu\nu}
-3\partial^2_{\nu\beta}D_{\mu\alpha}
+3\partial^3_{\mu\nu\beta}E_{\alpha}
\Bigr\}
\nonumber
\\
&&{}
+\dot{x}^\mu\dddot{x}^\nu\Bigl\{
\partial^2_{\alpha\beta}D_{\mu\nu}
-3\partial^2_{\mu\beta}D_{\nu\alpha}-\partial^2_{\nu\beta}D_{\mu\alpha}-\partial^2_{\mu\beta}D_{\alpha\nu}
\\
&&\hspace{17.5ex}{}
+2\partial_{\beta}B_{\mu\nu\alpha}-2\partial_{\beta}B_{\alpha\mu\nu}
+4\partial^2_{\mu\beta}C_{\nu\alpha}
+4\partial^3_{\mu\nu\beta}E_{\alpha}
\Bigr\}
\nonumber
\\
&&{}
+\ddddot{x}^\mu\Bigl\{
\partial^2_{\alpha\beta}E_{\mu}+\partial^2_{\mu\beta}E_{\alpha}+2\partial_{\beta}C_{\mu\alpha}
-\partial_{\beta}D_{\mu\alpha}-\partial_{\beta}D_{\alpha\mu}
\Bigr\}
\\
V_{(1)\alpha\beta}&=&
\dot{x}^\mu\dot{x}^\nu\dot{x}^\rho\Bigl\{
4\partial_{\alpha}A_{\mu\nu\rho\beta}-4\partial_{\beta}A_{\mu\nu\rho\alpha}-12\partial_{\rho}A_{\mu\nu\alpha\beta}
\\
&&\hspace{17.5ex}{}
+2\partial^2_{\nu\rho}B_{\mu\beta\alpha}+2\partial^2_{\beta\rho}B_{\mu\nu\alpha}
-\partial^3_{\mu\nu\rho}D_{\beta\alpha}-3\partial^3_{\beta\nu\rho}D_{\mu\alpha}+4\partial^4_{\beta\mu\nu\rho}E_{\alpha}
\Bigr\}
\\
&&{}
+\dot{x}^\mu\ddot{x}^\rho\Bigl\{
2\partial_{\alpha}B_{\mu\beta\rho}
+2\partial_{\rho}B_{\mu\beta\alpha}+4\partial_{\beta}B_{\rho\mu\alpha}+4\partial_{\mu}B_{\rho\beta\alpha}
-2\partial_{\beta}B_{\mu\alpha\rho}-2\partial_{\mu}B_{\alpha\beta\rho}
-24A_{\mu\rho\alpha\beta}
\\
&&\hspace{17.5ex}{}
+4\partial^2_{\mu\beta}C_{\rho\alpha}-6\partial^2_{\mu\beta}D_{\rho\alpha}
-3\partial^2_{\beta\rho}D_{\mu\alpha}-3\partial^2_{\mu\rho}D_{\beta\alpha}
+12\partial^3_{\beta\mu\rho}E_{\alpha}
\Bigr\}
\\
&&{}
+\dddot{x}^\mu\Bigl\{
\partial_{\alpha}D_{\beta\mu}
-3\partial_{\beta}D_{\mu\alpha}-\partial_{\mu}D_{\beta\alpha}-\partial_{\beta}D_{\alpha\mu}
+2B_{\mu\beta\alpha}-2B_{\alpha\beta\mu}
+4\partial_{\beta}C_{\mu\alpha}+4\partial^2_{\beta\mu}E_{\alpha}
\Bigr\}
\\
V_{(2)\alpha\beta}&=&
\dot{x}^\mu\dot{x}^\nu\Bigl\{
\partial_{\alpha}B_{\mu\nu\beta}+\partial_{\beta}B_{\mu\nu\alpha}
+4\partial_{\nu}B_{\mu\beta\alpha}-2\partial_{\nu}B_{\mu\alpha\beta}
\\*
&&\hspace{17.5ex}{}
-12A_{\mu\nu\alpha\beta}
+2\partial^2_{\mu\nu}C_{\alpha\beta}
-3\partial^2_{\mu\nu}D_{\beta\alpha}-3\partial^2_{\nu\beta}D_{\mu\alpha}+6\partial^3_{\beta\mu\nu}E_{\alpha}
\Bigr\}
\\
&&{}
+\ddot{x}^\mu\Bigl\{
2\partial_{\alpha}C_{\mu\beta}+2\partial_{\beta}C_{\mu\alpha}+2\partial_{\mu}C_{\alpha\beta}
+4B_{\mu\beta\alpha}-2B_{\mu\alpha\beta}-2B_{\alpha\beta\mu}
\\
&&\hspace{17.5ex}{}
-3\partial_{\beta}D_{\mu\alpha}
-3\partial_{\mu}D_{\beta\alpha}
+6\partial^2_{\beta\mu}E_{\alpha}
\Bigr\}
\\
V_{(3)\alpha\beta}&=&
\dot{x}^\mu\Bigl\{
\partial_{\alpha}D_{\mu\beta}
-3\partial_{\mu}D_{\beta\alpha}-\partial_{\beta}D_{\mu\alpha}-\partial_{\mu}D_{\alpha\beta}
+2B_{\mu\beta\alpha}-2B_{\mu\alpha\beta}+4\partial_{\mu}C_{\alpha\beta}+4\partial^2_{\beta\mu}E_{\alpha}
\Bigr\}
\\
V_{(4)\alpha\beta}&=&
\partial_{\alpha}E_{\beta}+\partial_{\beta}E_{\alpha}+2C_{\alpha\beta}-D_{\beta\alpha}-D_{\alpha\beta}
\end{eqnarray*}
Here the function $\delta_{\partial M}(z)$ is defined as follows
\begin{eqnarray}
\frac{\delta x^\mu(t)}{\delta x^\nu(z')}=\delta(t-t')
\delta_{\partial M}(z')\delta^\mu_\nu 
\end{eqnarray}
and if we use $(t,y)$ coordinates $\delta_{\partial M}(z)=2\delta(y)$.

The divergences of (\ref{3.5}) are contained in the expression
\begin{eqnarray}
\frac{\alpha'^{3/2}}{2} \sum_{k=0}^4 \int\!\!\!dtV_{(k)\alpha\beta}(t)
\left(\frac{d^k}{dt^k} G^{\beta\alpha}(t,0;t',0)\right)\bigg|_{t'\to t}=
-\frac{\mu^\epsilon}{\epsilon}\frac{\alpha'^{3/2}}{2\pi}
\int dt V_{(0)\alpha\beta}(t)\eta^{\alpha\beta},
\label{3.12}
\end{eqnarray}
where we used that the divergences of Green function in coincident points
in the framework of dimensional renormalization are \cite{30,31}:
\begin{eqnarray}
\nonumber
G^{\mu\nu}(t,0;t',0)\Big|^{\text{div}}_{t'\to t}
&=&
- \frac{\mu^\epsilon}{\pi\epsilon} \eta^{\mu\nu} ,
\\
\label{3.13}
\frac{d^k}{dt^k}G^{\mu\nu}(t,0;t',0)\Big|^{\text{div}}_{t'\to t}&=&
=0,
\qquad
k>0.
\end{eqnarray}
Omitting in (\ref{3.12}) total derivatives we arrive at the following
one-loop effective action:
\begin{eqnarray}
\nonumber
\Gamma^{(1)}
&=&
\frac{\alpha'^{3/2}\mu^\epsilon}{2\pi}\int\limits_{\partial M}
\!\!\!dte\left(\dot{x}^\mu\dot{x}^\nu\dot{x}^\rho\dot{x}^\sigma(A_{\mu\nu\rho\sigma}
-\frac{1}{\epsilon}\Box A_{\mu\nu\rho\sigma})+\right.
\\
\label{3.14}
&&{}
+\dot{x}^\mu\dot{x}^\nu\ddot{x}^\rho
(B_{\mu\nu\rho}-\frac{1}{\epsilon}\Box B_{\mu\nu\rho})
+\ddot{x}^\mu\ddot{x}^\nu
(C_{\mu\nu}-\frac{1}{\epsilon}\Box C_{\mu\nu})
\\
\nonumber
&&
+\dot{x}^\mu\dddot{x}^\nu
(D_{\mu\nu}-\frac{1}{\epsilon}\Box D_{\mu\nu})
+\ddddot{x}^\mu
(E_{\mu}-\frac{1}{\epsilon}\Box E_{\mu})
+(fin),
\\
\nonumber
& &\Box\equiv\eta^{\alpha\beta}\partial_\alpha\partial_\beta,
\end{eqnarray}
where $(fin)$ stands for a finite part of the one-loop correction.

To cancel the divergences in (\ref{3.14}) renormalization of all the
background fields should be of the form
\begin{eqnarray}
\label{3.15}
\stackrel{\circ}{\Phi} = \mu^{\epsilon}(\Phi + \frac{1}{\epsilon} \Box \Phi )
\end{eqnarray}
where $\Phi=(A_{\mu\nu\rho\sigma},B_{\mu\nu\rho},C_{\mu\nu},D_{\mu\nu},E_\mu)$
and $\circ$ denotes the bare background fields.

After the renormalization of the background fields, we should renormalize the trace of the energy-momentum tensor to impose the condition of Weyl invariance of the theory at the quantum level.

\section{\bf Renormalization of the trace of the energy-momentum tensor}\setcounter{equation}{0}

In classical theory trace of the energy-momentum tensor for the theory
(\ref{300}) on the $2+\epsilon$-dimensional world sheet is
\begin{eqnarray}
T(z) &=& \frac{g_{ab}(z)}{\sqrt{g(z)}}\; \frac{\delta S}{\delta g_{ab}(z)}\Bigm|_{g_{ab}\to\delta_{ab}} 
=
\frac{\epsilon}{8\pi}g^{ab}(z)\partial_a x^\mu\partial_b x^\nu\delta_{\mu\nu}
\nonumber
\\
&&{}
+\frac{\alpha '^{3/2}(1+\epsilon)}{4\pi}\;\mathcal{A}_{\mu\nu\rho\sigma}\;\dot{x}^\mu\dot{x}^\nu\dot{x}^\rho\dot{x}^\sigma
\;\delta_{\partial M}(z)
+\frac{\alpha '^{3/2}(1+\epsilon)}{4\pi}\;\mathcal{B}_{\mu\nu\rho}\;\dot{x}^\mu\dot{x}^\nu\ddot{x}^\rho
\;\delta_{\partial M}(z)
\nonumber
\\
&&{}
-\frac{\alpha '^{3/2}(1+\epsilon)}{4\pi}\;\mathcal{C}_{\mu\nu}\;\ddot{x}^\mu\ddot{x}^\nu
\;\delta_{\partial M}(z)
+\frac{\alpha '^{3/2}(1+\epsilon)}{2\pi}\;\mathcal{C}_{\mu\nu}\;\dot{x}^\mu\dddot{x}^\nu 
\;\delta_{\partial M}(z)
\end{eqnarray}
where
\begin{eqnarray*}
\mathcal{A}_{\mu\nu\rho\sigma}&=&
-3 A_{\mu\nu\rho\sigma}+\partial_{(\sigma}B_{\mu\nu\rho)}
-\partial^2_{(\rho\sigma}D_{\mu\nu)}+\partial^3_{(\nu\rho\sigma}E_{\mu)}
\\
\mathcal{B}_{\mu\nu\rho}&=&
-2B_{\mu\nu\rho}+2B_{\rho\mu\nu}
+2\partial_{\nu}C_{\rho\mu}+\partial_{\nu}D_{\mu\rho}
-\partial_{\rho}D_{\mu\nu}-2\partial_{\nu}D_{\rho\mu}
+3\partial^2_{\nu\rho}E_{\mu}-\partial^2_{\mu\nu}E_{\rho}
\\
\mathcal{C}_{\mu\nu}&=&
C_{\mu\nu}-D_{(\mu\nu)}+\partial_{(\mu}E_{\nu)}
\end{eqnarray*}

To calculate the trace of the energy-momentum tensor in quantum theory we
should define renormalized values for composite operators. Consider, for
example, the vacuum average for one of these operators:
\begin{eqnarray}
\nonumber
\langle\mathcal{C}_{\mu\nu}(x)\ddot{x}^\mu\ddot{x}^\nu\rangle=\int Dxe^{-S[x]}\mathcal{C}_{\mu\nu}(x)
\ddot{x}^\mu\ddot{x}^\nu
\biggm/
\int Dxe^{-S[x]}.
\end{eqnarray}
Making the shift $x^\mu=\bar{x}^\mu+\zeta^\mu$ in the functional integral
($\bar{x}^\mu$ are solutions of the classical equations of motion) and
using (\ref{3.13}) we get in the linear approximation:
\begin{eqnarray}
\langle\mathcal{C}_{\mu\nu}(x)\ddot{x}^\mu\ddot{x}^\nu\rangle=\mu^\epsilon\ddot{\bar{x}}^\mu
\ddot{\bar{x}}^\nu(\mathcal{C}_{\mu\nu}(\bar{x})-\frac{1}{\epsilon}\Box
\mathcal{C}_{\mu\nu}(\bar{x}))+(fin).
\end{eqnarray}
Renormalized operators should have a finite average value
\begin{eqnarray}
\langle[\mathcal{C}_{\mu\nu}(x)\ddot{x}^\mu\ddot{x}^\nu]\rangle=\mathcal{C}_{\mu\nu}(\bar{x})
\ddot{\bar{x}}^\mu\ddot{\bar{x}}^\nu+ (fin)
\end{eqnarray}
hence
\begin{eqnarray}
(\mathcal{C}_{\mu\nu}(x)\ddot{x}^\mu\ddot{x}^\nu)_\circ=\mu^\epsilon[\ddot{x}^\mu
\ddot{x}^\nu(\mathcal{C}_{\mu\nu}(x)-\frac{1}{\epsilon}\Box\mathcal{C}_{\mu\nu}(x))]+(fin),
\end{eqnarray}
or, using (\ref{3.15}),
\begin{eqnarray}
(\stackrel{\!\!\!\!\!\circ}{\mathcal{C}_{\mu\nu}} \ddot{x}^\mu \ddot{x}^\nu)_\circ =
[\mathcal{C}_{\mu\nu} \ddot{x}^\mu \ddot{x}^\nu ]
\end{eqnarray}
In the same way one can get
\begin{eqnarray*}
(\stackrel{\!\!\circ}{\mathcal{A}}_{\mu\nu\rho\sigma}\;\dot{x}^\mu\dot{x}^\nu\dot{x}^\rho\dot{x}^\sigma)_\circ
&=&
[ \mathcal{A}_{\mu\nu\rho\sigma}\;\dot{x}^\mu\dot{x}^\nu\dot{x}^\rho\dot{x}^\sigma],
\\
(\stackrel{\circ}{\mathcal{B}}_{\mu\nu\rho}\;\dot{x}^\mu\dot{x}^\nu\ddot{x}^\rho)_\circ
&=&
[ \mathcal{B}_{\mu\nu\rho}\;\dot{x}^\mu\dot{x}^\nu\ddot{x}^\rho ]
\\
(\stackrel{\!\!\!\!\!\circ}{\mathcal{C}_{\mu\nu}} \dot{x}^\mu \dddot{x}^\nu)_\circ 
&=&
[\mathcal{C}_{\mu\nu} \dot{x}^\mu \dddot{x}^\nu ]
\end{eqnarray*}

Similar but more tedious calculation give the following renormalization of
the operator $g^{ab}(z)\partial_ax^\mu\partial_bx^\nu\delta_{\mu\nu}$:
\begin{eqnarray*}
(g^{ab}\partial_ax^\mu\partial_bx^\nu\delta_{\mu\nu})_\circ &=&
[g^{ab}\partial_ax^\mu\partial_bx^\nu\delta_{\mu\nu}]+
\\
&&\hspace{-15ex}{}
+\alpha'^{3/2}\frac{\mu^\epsilon}{\epsilon}\Bigl[
2V_{(0)}-\frac{d}{edt}V_{(1)}
+\frac{1}{2}\frac{d^2}{(edt)^2}V_{(2)}
-\frac{1}{4}\frac{d^3}{(edt)^3}V_{(3)}
+\frac{1}{8}\frac{d^4}{(edt)^4}V_{(4)}
\Bigr]\delta_{\partial M}(z) 
\\
&&\hspace{-15ex}{}
+\alpha'^{3/2}\frac{\mu^\epsilon}{\epsilon}\Bigl[
-\frac{1}{2}V_{(2)}
+\frac{3}{4}\frac{d}{edt}V_{(3)}
-\frac{3}{4}\frac{d^2}{(edt)^2}V_{(4)}
\Bigr]\delta''_{\partial M}(z) 
\\
&&\hspace{-15ex}{}
+\alpha'^{3/2}\frac{\mu^\epsilon}{\epsilon}\Bigl[\frac{1}{8}V_{(4)}\Bigr]\delta''''_{\partial M}(z) 
\\[1em]
& &V_{(k)}\equiv V_{(k)\mu\nu}\delta^{\mu\nu},
\qquad
\delta''_{\partial M}(z)=\partial_y^2\delta_{\partial M}(z)
\qquad
\delta''''_{\partial M}(z)=\partial_y^4\delta_{\partial M}(z)
\end{eqnarray*}
As a result, the renormalized operator of the energy-momentum tensor trace
has the form
\begin{eqnarray}
[T]&=&
\frac{\alpha'^{3/2}}{8\pi}\Bigl[
2V_{(0)}-\frac{d}{edt}V_{(1)}
+\frac{1}{2}\frac{d^2}{(edt)^2}V_{(2)}
-\frac{1}{4}\frac{d^3}{(edt)^3}V_{(3)}
+\frac{1}{8}\frac{d^4}{(edt)^4}V_{(4)}
\Bigr]\delta_{\partial M}(z) 
\nonumber
\\
&&{}
+\frac{\alpha'^{3/2}}{8\pi}\Bigl[
-\frac{1}{2}V_{(2)}
+\frac{3}{4}\frac{d}{edt}V_{(3)}
-\frac{3}{4}\frac{d^2}{(edt)^2}V_{(4)}
\Bigr]\delta''_{\partial M}(z) 
\nonumber
\\
&&{}
+\frac{\alpha'^{3/2}}{8\pi}\Bigl[\frac{1}{8}V_{(4)}\Bigr]\delta''''_{\partial M}(z) 
\nonumber
\\
&&{}
+\frac{\alpha '^{3/2}}{4\pi}
\;\Bigl[\mathcal{A}_{\mu\nu\rho\sigma}\;\dot{x}^\mu\dot{x}^\nu\dot{x}^\rho\dot{x}^\sigma\Bigr]
\;\delta_{\partial M}(z)
+\frac{\alpha '^{3/2}}{4\pi}\;\Bigl[\mathcal{B}_{\mu\nu\rho}\;\dot{x}^\mu\dot{x}^\nu\ddot{x}^\rho\Bigr]
\;\delta_{\partial M}(z)
\nonumber
\\
&&{}
-\frac{\alpha '^{3/2}}{4\pi}\;\Bigl[\mathcal{C}_{\mu\nu}\;\ddot{x}^\mu\ddot{x}^\nu\Bigr]
\;\delta_{\partial M}(z)
+\frac{\alpha '^{3/2}}{2\pi}\;\Bigl[\mathcal{C}_{\mu\nu}\;\dot{x}^\mu\dddot{x}^\nu\Bigr] 
\;\delta_{\partial M}(z)
\label{4.10}
\end{eqnarray}
From the requirement of quantum Weyl invariance it follows that all the coefficients at the linear independent operators in
(\ref{4.10}) should be equal to zero.
Let us analyze the obtained equations.

\section{\bf Analysis of the equations of motions}\setcounter{equation}{0}

In this section, we will simplify the equation on the background fields, which follow from the vanishing of the renormalized trace of the energy-momentum tensor(\ref{4.10}).

First we consider coefficient at $\delta''''_{\partial M}(z)$. It gives equation $V_{(4 )}=0$.
Next we note that $V_{(3)}=2\frac{d}{edt}V_{(4)}$ and therefore coefficient at  $\delta''_{\partial M}(z)$ gives equation $V_{(2 )}=0$.

To simplify further analysis we fix the symmetry (\ref{3.2}) and choose the following gauge conditions
\begin{eqnarray}
B_{(\mu\nu\rho)}=0,
\qquad
D_{\mu\nu}=0,
\qquad
E_\mu=0.
\label{GCond}
\end{eqnarray}
Under these conditions the equations of motion take the form
\begin{eqnarray}
V_{(4)}=0&\quad\Rightarrow\quad&C^\alpha{}_{\alpha}=0
\label{V4}
\\
V_{(2)}=0&\Rightarrow&
\dot{x}^\mu\dot{x}^\nu\Bigl\{
2\partial^{\alpha}B_{\mu\nu\alpha}+2\partial_{\nu}B_{\mu\alpha}{}^{\alpha}
-12A_{\mu\nu\alpha}{}^{\alpha}
+2\partial^2_{\mu\nu}C^{\alpha}{}_{\alpha}
\Bigr\}
=0
\label{V21}
\\
&&{}
\ddot{x}^\mu\Bigl\{
4\partial^{\alpha}C_{\mu\alpha}+2\partial_{\mu}C^{\alpha}{}_{\alpha}
+2B_{\mu\alpha}{}^{\alpha}-2B^{\alpha}{}_{\alpha\mu}
\Bigr\}
=0
\label{V22}
\end{eqnarray}
and the part of renormalized trace of EMT which proportional to $\delta_{\partial M}(z)$ gives
\begin{eqnarray}
&&
\dot{x}^\mu\dot{x}^\nu\dot{x}^\rho\dot{x}^\sigma\Bigl\{
\Box A_{\mu\nu\rho\sigma}-3 A_{\mu\nu\rho\sigma}
-4\partial^2_{\sigma\alpha}A_{\mu\nu\rho}{}^{\alpha}
+6\partial^2_{\rho\sigma}A_{\mu\nu\alpha}{}^{\alpha}
-\partial^3_{\nu\rho\sigma}B_{\mu\alpha}{}^{\alpha}
\Bigr\}
=0
\label{dddd}
\\
&&{}
\dot{x}^\mu\dot{x}^\nu\ddot{x}^\rho\Bigl\{
\Box B_{\mu\nu\rho}-3B_{\mu\nu\rho}
-2\partial^2_{\nu\alpha}B^{\alpha}{}_{\mu\rho}
-3\partial^2_{\nu\rho}B_{\mu\alpha}{}^{\alpha}
-3\partial^2_{\mu\nu}B_{\rho\alpha}{}^{\alpha}
+\partial^2_{\mu\nu}B^{\alpha}{}_{\alpha\rho}
\nonumber
\\
&&\hspace{17.5ex}{}
-12\partial^{\alpha}A_{\mu\nu\rho\alpha}
+6\partial_{\rho}A_{\mu\nu\alpha}{}^{\alpha}
+24\partial_{\nu}A_{\mu\rho\alpha}{}^{\alpha}
+2\partial_{\nu}C_{\rho\mu}
\Bigr\}
=0
\label{dd2d}
\\
&&{}
\ddot{x}^\mu\ddot{x}^\nu\Bigl\{
\Box C_{\mu\nu}
-C_{\mu\nu}
-2\partial^{\alpha}B_{\alpha\mu\nu}
-3\partial_{\nu}B_{\mu\alpha}{}^{\alpha}
+\partial_{\mu}B^{\alpha}{}_{\alpha\nu}
+12A_{\mu\nu\alpha}{}^{\alpha}
\Bigr\}
\label{2d2d}
=0
\\
&&{}
\dot{x}^\mu\dddot{x}^\nu \Bigl\{
12A_{\mu\nu\alpha}{}^{\alpha}
-2\partial^{\alpha}B_{\alpha\mu\nu}
-\partial_{\nu}B_{\mu\alpha}{}^{\alpha}
-3\partial_{\mu}B_{\nu\alpha}{}^{\alpha}
+2\partial_{\mu}B^{\alpha}{}_{\alpha\nu}
+2C_{\mu\nu}
\Bigr\}
=0
\label{d3d}
\\
&&{}
\ddddot{x}^\mu\Bigl\{
B^{\alpha}{}_{\alpha\mu}-B_{\mu\alpha}{}^{\alpha}
\Bigr\}
=0
\label{4dotx}
\end{eqnarray}

First from \eqref{4dotx} and the gauge condition \eqref{GCond} 
we find that $B^{\alpha}{}_{\alpha\mu}=B_{\mu\alpha}{}^{\alpha}=0$
and from \eqref{V4} $C^\alpha{}_{\alpha}=0$.
Next equations \eqref{V21} and \eqref{V22} give $\partial^{\alpha}B_{\mu\nu\alpha}=6A_{\mu\nu\alpha}{}^{\alpha}$
and $\partial^{\alpha}C_{\mu\alpha}=0$, respectively,
and subtracting \eqref{d3d} from \eqref{2d2d} we find $\Box C_{\mu\nu}-C_{\mu\nu}=0$.
Analogusly, using \eqref{d3d} equation \eqref{dd2d} reduces to
\begin{eqnarray}
&&{}
\dot{x}^\mu\dot{x}^\nu\ddot{x}^\rho\Bigl\{
\Box B_{\mu\nu\rho}-3B_{\mu\nu\rho}
-12\partial^{\alpha}A_{\mu\nu\rho\alpha}
+18\partial_{(\rho}A_{\mu\nu)\alpha}{}^{\alpha}
\Bigr\}
=0.
\label{dd2d+}
\end{eqnarray}
Extracting from this equation completely symmetric part we find 
$3\partial_{(\rho}A_{\mu\nu)\alpha\alpha}=2\partial_{\alpha}A_{\mu\nu\rho\alpha}$
and using this relation we get from \eqref{dddd} and \eqref{dd2d+}, respectively,
\begin{eqnarray*}
\Box A_{\mu\nu\rho\sigma}-3 A_{\mu\nu\rho\sigma}=0,
&\qquad&
\Box B_{\mu\nu\rho}-3B_{\mu\nu\rho}=0.
\end{eqnarray*}
Finally, dividing \eqref{d3d} into symmetric and antisymmetric parts one finds
\begin{eqnarray*}
6A_{\mu\nu\alpha}{}^{\alpha}+C_{\mu\nu}=\partial^{\alpha}B_{\alpha(\mu\nu)},
&\qquad&
\partial^{\alpha}B_{\alpha[\mu\nu]}=0.
\end{eqnarray*}

Let us summarize the obtained equations. 
Returning to the dimensional string coordinates $x^\mu\to\alpha'^{-1/2}x^\mu$ and denoting $m^2=1/\alpha'$ one can rewrite these equations as
\begin{subequations}
\begin{eqnarray}
&&
\Box A_{\mu\nu\rho\sigma}
-3m^2 A_{\mu\nu\rho\sigma}
=0
\\
&&
3\partial_{(\rho}A_{\mu\nu)\alpha}{}^{\alpha}=2\partial^{\alpha}A_{\mu\nu\rho\alpha}
\\
&&
\Box B_{\mu\nu\rho}-3m^2B_{\mu\nu\rho}
=0
\qquad
B^{\alpha}{}_{\alpha\mu}=B_{\mu\alpha}{}^{\alpha}=0
\\
&&
\Box C_{\mu\nu}-3m^2C_{\mu\nu}=0
\qquad
\partial^{\alpha}C_{\mu\alpha}=0
\qquad
C^\alpha{}_{\alpha}=0
\label{eqnsC}
\\
&&
9A_{\mu\nu\alpha}{}^{\alpha}+C_{\mu\nu}=0
\label{eqnsC+}
\\
&&
6mA_{\mu\nu\alpha}{}^{\alpha}=\partial^{\alpha}B_{\mu\nu\alpha}
\qquad
\partial^{\alpha}B_{\alpha\mu\nu}=\partial^{\alpha}B_{\alpha\nu\mu}
\end{eqnarray}
\label{eqns}
\end{subequations}
Let us compare equations\eqref{eqns} with the equations of the third massive level of the open string spectrum.
As is known the third massive level of the open string spectrum consists of spin 4, spin 2, spin 0 fields,
and rang-3 tensor field with mixed index symmetry.
We see that we do not have any scalar field and we can not make it out of the other fields since it will be zero if we use \eqref{eqns}.
We suppose that the scalar field should appear if we will consider curved world sheet.
In this case, we will have several scalar fields, one of which may be a scalar field of the string spectrum.
We also note that in case of a curved world sheet there is no room  for additional spin 4 field and for additional rang-3 field with mixed index symmetry.

Let us check whether the fields $A_{\mu\nu\rho\sigma}$, $C_{\mu\nu}$, $B_{\mu\nu\rho}$ describe  spin 4, spin 2, and rang-3 tensor field with mixed index symmetry, respectively.
Field $C_{\mu\nu}$ satisfies all the conditions to be spin 2 field \eqref{eqnsC}, but it is restricted one more equation  \eqref{eqnsC+}, that is this field is not independent. 
In case of a curved world sheet equation \eqref{eqnsC+} can be modified only by additional background fields. 
Therefore we can assume that the field $C_{\mu\nu}$ can not be a field from the string spectrum and should become zero in this case.
As for the remained two fields $A_{\mu\nu\rho\sigma}$ and $B_{\mu\nu\rho}$, we can see from \eqref{eqns} they indeed describe the spin 4 field and the field with mixed index symmetry if we suppose $C_{\mu\nu}=0$. 

Thus, we have shown that in order to reproduce a full set of equations defining the third massive level of the open string spectra one should consider the case of cureved world sheet of the string. The case of flat world sheet gives us only a part of the equations on the background fields.

\section{\bf Summary}
We have considered sigma-model description of an open string interacting with massive fields of the fourth (third massive) level and have constructed the one-loop renormalization of the background fields
and the energy-momentum tensor in the linear approximation in the massive background fields. 
The equations of motion for the background fields have been obtained from the
requirement of quantum Weyl invariance.
It has been shown that the obtained equations are consistent with the structure of the open string
spectrum at the third massive level.
It would be interesting to repeat our analysis to consider the case when the world sheet of the string is not flat
and/or to add the graviton field to the sigma-model action of interacting open string.

\section*{\bf Acknowledgements}
The author are grateful to I.L. Buchbinder for setting the problem and its discussion,
and to RFBR grant project No. 15-02-03594 for partial support.
The research was also supported in parts by Russian Ministry of Education and
Science, project No. 3.1386.2017.

\end{document}